\documentclass[twocolumn,aps,prb,superscriptaddress]{revtex4-1}
\usepackage{bm}
\usepackage{amsmath,amsthm,amssymb}
\usepackage{color}
\usepackage{newtxmath}
\usepackage{array,booktabs}
\usepackage[dvipdfmx]{graphicx}
\usepackage{here}

\begin{document}

\title{Unconventional gate voltage dependence of the charge conductance caused by spin-splitting Fermi surface by Rashba-type spin-orbit coupling}

\author{Daisuke Oshima}
\affiliation{Department of Applied Physics, Nagoya University, Nagoya, 464-8603, Japan}

\author{Katsuhisa Taguchi}
\affiliation{Yukawa Institute for Theoretical Physics, Kyoto University, Kyoto, 606-8502, Japan}

\author{Yukio Tanaka}
\affiliation{Department of Applied Physics, Nagoya University, Nagoya, 464-8603, Japan}

\begin{abstract}
  We calculate the gate voltage ($V_g$) dependence of charge conductance in a normal metal (NM)/two dimensional electron gas (2DEG) junction, where Rashba spin-orbit coupling and ferromagnetism exist in the 2DEG.
  We call this 2DEG as the ferromagnetic Rashba metal (FRM) and the chemical potential of the FRM is controlled by $V_{g}$.
  We clarify the physical origin of the unconventional $V_{g}$ dependence of charge conductance in the NM/FRM junction found in our previous work [J. Phys. Soc. Jpn. {\bf 87}, 034710 (2018)], in which the charge conductance increases with $V_{g}$, although the number of carries in FRM decreases.
  We calculate the momentum-resolved charge conductance.
  It is clarified that the origin of the unconventional $V_{g}$ dependence is due to the non-monotonic change in the size of the inner Fermi surface in FRM as a function of $V_{g}$.
  \end{abstract}

\maketitle

\section{Introduction}
\label{sec:I}
Spin-orbit coupling (SOC), which leads to spin-momentum locking and band-splitting, plays an important role in solid state physics.
In particular, Rashba-type spin-orbit coupling (RSOC)\cite{Rashba60} can be controlled by an electric field\cite{Datta90,Nitta97}, and RSOC triggers spin-dependent transport.
Thus, RSOC is a central issue of spintronics, and several unconventional charge transports have been reported by RSOC\cite{Molenkamp01,Jiang03,Cai08,Jiang10}.
In the presence of RSOC in a two-dimensional electron gas (2DEG), the energy band of 2DEG splits and there are two Fermi surfaces with different directions of the spin helicity.
As a result, RSOC affects the charge transport of the 2DEG.
Thus, the gate voltage ($V_{g}$) dependence (i.e., Fermi level dependence) of the charge conductance of normal metal (NM)/2DEG junctions has been studied\cite{Streda03,Srisongmuang08,Tang12,Jantayod13,Rainis14,Jantayod15,Cayao15,Tang17,Oshima18}.

\begin{figure}[htbp]
  \centering
  \includegraphics[width = 80mm]{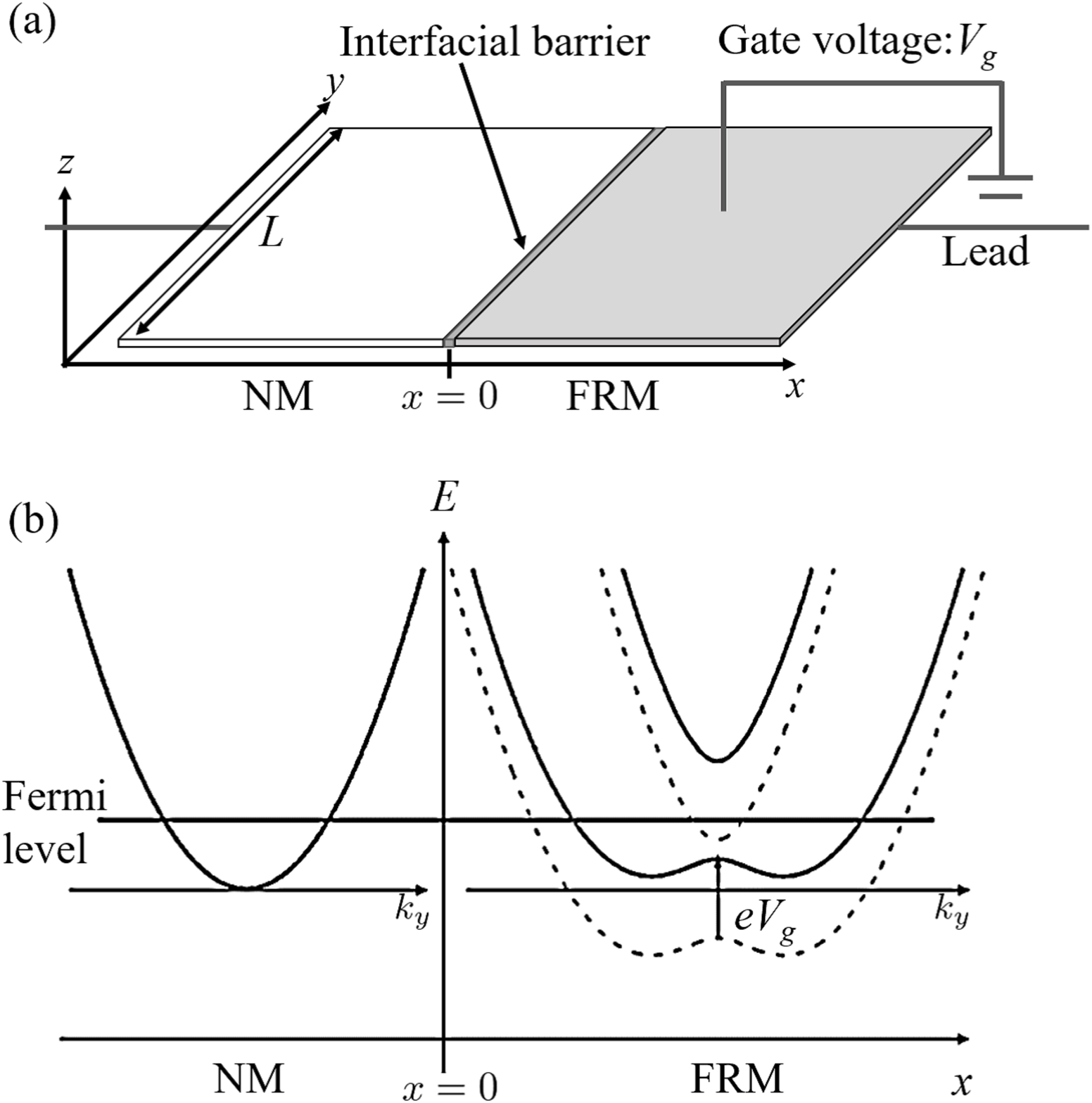}
  \caption{(a) Schematic of a two-dimensional NM/FRM junction, where the FRM is a 2DEG with RSOC and ferromagnetism.
  Gate voltage ($V_g$) is applied on the FRM.
  $L$ is the width of the junction along the $y$ direction.
  (b) Energy dispersion of the NM (left panel) and FRM (right panel) in the junction. The band of the FRM is shifted by $V_g$, in which dashed and solid lines indicate the energy band before and after applying $V_g$, respectively.}
  \label{junctionmodel}
  \end{figure}
%
In an NM/2DEG junction, the position of the band bottom of the 2DEG can be controlled by gate voltage $V_g$, applied to the 2DEG.
The carrier number and the magnitude of group velocity of the 2DEG are basically small when compared to those in an NM.
We assume that the position of the band bottom moves upward and the number of carriers in the 2DEG decreases by $V_g$.
If we ignore the RSOC in the 2DEG, the charge conductance $\mathcal{G}$ in the NM/2DEG junction shows $\partial\mathcal{G}/\partial(eV_g)<0$, because the magnitude of the Fermi momentum in the 2DEG monotonically decreases with increasing $V_{g}$.
On the other hand, it has been clarified that in the presence of RSOC, $\mathcal{G}$ in an NM/ferromagnetic Rashba metal (FRM) junction (see Fig. \ref{junctionmodel}) is such that $\partial\mathcal{G}/\partial(eV_g)>0$, unlike in conventional charge conductance \cite{Rainis14,Cayao15,Oshima18}.
Here, the FRM has RSOC and ferromagnetism (detail is shown in Sec. \ref{sec:II}), which could cause the unconventional property of $\mathcal{G}$.
However, it has not been clarified why $\partial\mathcal{G}/\partial(eV_g)>0$ is satisfied.

The aim of this study is to clarify the physical origin of the sign change of $\partial \mathcal{G}/\partial (eV_g)$ from negative to positive
by RSOC.
We clarify that the sign change appears only for strong RSOC.
In addition, through momentum-resolved conductance, we show that the sign change stems from the contribution of the inner Fermi surface caused by RSOC.
We clarify that its contribution to $\mathcal{G}$ is sensitive to the interfacial barrier and vanishes for large magnitudes of the interfacial barrier.

The organization of this paper is as follows.
In Sect. \ref{sec:II}, we show a model of the NM/FRM junction and the method of calculating charge conductance $\mathcal{G}$.
In Sect. \ref{sec:III}, we show the obtained results and discuss their physical meaning.
In Sect. \ref{sec:V}, we summarize the results.

\section{Model and method}
\label{sec:II}
\subsection{Model of FRM}
First, we introduce the FRM system.
The model Hamiltonian of FRM is described as\cite{Streda03,Rainis14,Fukumoto15,Cayao15,Oshima18}
\begin{align}
H_{\rm FRM} = \frac{\hbar^2k^2}{2m_R}+{H}_{{\rm RSOC}}-M\sigma_z,
\end{align}
where $\hbar^2k^2/(2m_{R})$ denotes the kinetic energy, $k^2=k^2_x+k^2_y$, and $m_R$ is the effective mass in the FRM.
$H_{\rm RSOC}$ denotes the RSOC\cite{Ganichev14,Kohda17}:
\begin{align}
H_{\rm RSOC}=\alpha(k_x\sigma_y-k_y\sigma_x),
\end{align}
where ${\bm \sigma}=(\sigma_x,\sigma_y,\sigma_z)$ are Pauli matrices in the spin space, and $\alpha$ is a coupling constant.
$-M\sigma_z$ denotes the exchange coupling of the magnetization.
Here, we assume perpendicular magnetization for simplicity.

\begin{figure*}[htbp]
  \centering
  \includegraphics[width = 180mm]{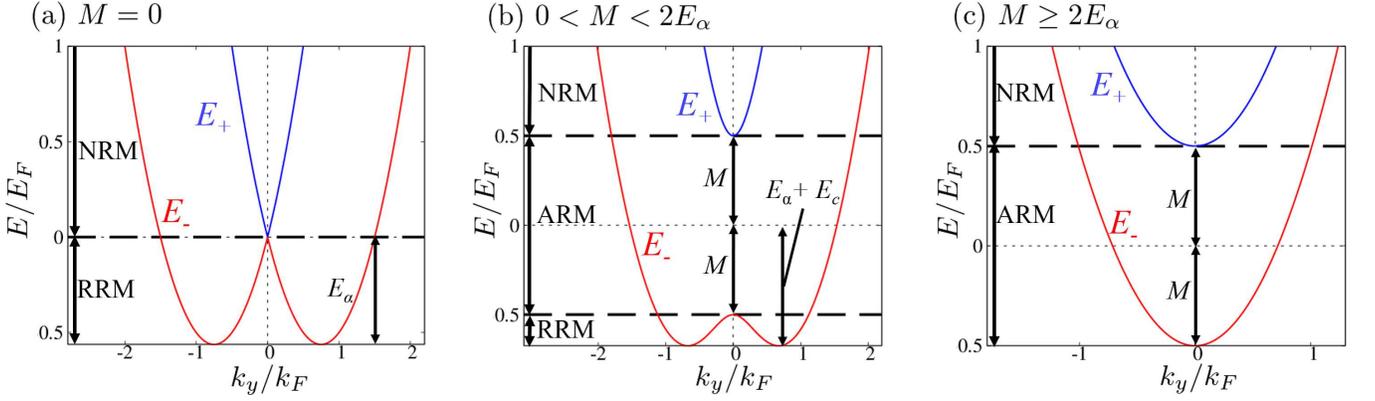}
  \caption{(Color online) Band structure of FRM.
  (a) $M=0$, (b) $0<M<2E_\alpha$, and (c) $M>2E_\alpha$, where $E_\alpha$ is the Rashba energy, and $M$ is the magnitude of the magnetization of the FRM.
  $E_c=M^2/(4E_\alpha)$.
  The blue and red lines correspond to the eigenvalues $E_+$ and $E_-$, respectively.
  Here, we set (a) $E_\alpha/E_F=0.55$, (b) $E_\alpha/E_F=0.55$, $M/E_F=0.5$, and (c) $E_\alpha/E_F=0.02$, $M/E_F=0.5$, where $E_F$ and $k_F$ are the Fermi energy and Fermi momentum in the NM, respectively.
  }
  \label{bandstructure}
\end{figure*}
\begin{figure}[htbp]
  \centering
  \includegraphics[width = 85mm]{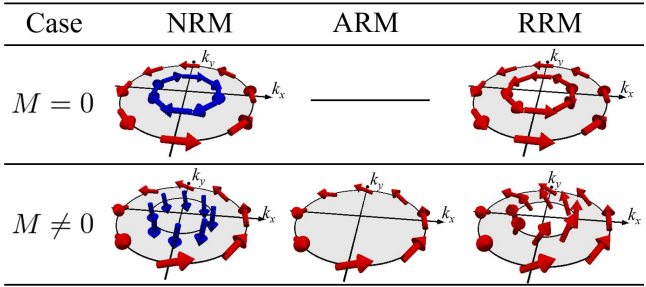}
  \caption{(Color online) Spin texture of the NRM, ARM, and RRM states in the momentum ($k_x$-$k_y$) space with $M=0$ and $M \neq 0$.
  Red and blue arrows denote the spin polarization of $E_-$ and $E_+$ bands, respectively.
  The shaded region shows the occupied state.}
  \label{spintexture}
\end{figure}
%
For $M=0$, there are two spin-split bands: $E_\pm=\frac{\hbar^2 k^2}{2m_R}\pm \alpha |k|$ [see Fig. \ref{bandstructure}(a)].
These bands cross at $E=0$ and have an annular minimum at $E=-E_\alpha$.
Here, $E_\alpha=m_R \alpha^2/(2\hbar^2)$ is the Rashba energy \cite{Srisongmuang08,Jantayod13,Cayao15,Ast07,Sablikov07,Ast08,Mathias10,Ishizaka11}.
For $E>0$, the Fermi level crosses $E_+$ and $E_-$ bands, and there are two Fermi surfaces with opposite spin direction (see Fig. \ref{spintexture}).
We call this state the normal Rashba metal (NRM) state.
For $-E_\alpha<E<0$, Fermi level crosses only $E_-$ band, but there are two Fermi surfaces with the same spin direction.
In this region, the shape of the occupied state in the momentum space is ring shape (see Fig. \ref{spintexture}).
Then, we name this state Rashba ring metal (RRM) state\cite{Oshima18}.

For $M\neq 0$, two bands $E_\pm=\frac{\hbar^2 k^2}{2m_R}\pm \sqrt{\alpha^2 k^2 +M^2}$ perfectly split.
Only one Fermi surface appears for $-M<E<M$ (see Fig. \ref{spintexture}).
We call this the anomalous Rashba metal (ARM) state\cite{Fukumoto15}.
Here, the shape of the band structure depends on whether $2E_\alpha >M$ is satisfied.
For $2E_\alpha >M$, the band structure has an annular minimum at $E=-E_\alpha-E_c$, with $E_c=M^2/(4E_\alpha)$ [see Fig. \ref{bandstructure}(b)].
There are NRM, ARM, and RRM states.
For $2E_\alpha \leq M$, the band structure has a point minimum at $E=-M$ [see Fig. \ref{bandstructure}(c)], and the RRM state does not appear.
It is noted that the spin direction on the Fermi surfaces leans to the $z$-direction because of the perpendicular magnetization.

\begin{table}[htbp]
  \centering
  \caption{Energy ($E$) dependence of the DOS of the FRM for the NRM, ARM, and RRM states\cite{Oshima18}.
  Here, we set $\epsilon=E+E_\alpha+E_c$, and $2\nu_e$ is the DOS of the 2DEG.
  The NRM, ARM, and RRM states are realized for $M<E$, $-M<E<M$, and $-E_\alpha-E_c<E<-M$, respectively. The DOS is defined as ${\rm Tr}[\bm{\rho}]$ with $\bm{\rho}(E)\equiv -1/\pi {\rm Im}\sum_k (E-H_{\rm FRM}+i\delta)^{-1}$. $\delta$ is an infinitesimal positive value.}
  \begin{tabular}{cccc}
  \hline
  Case& NRM & ARM & RRM \\
  \hline
  $M=0 $ & $2\nu_e$ & -&$2\nu_e\sqrt{\frac{E_\alpha}{\epsilon}}$\\
  $0<M<2E_\alpha$   & $2\nu_e$ & $\nu_e \left[ 1+ \sqrt{\frac{E_\alpha}{\epsilon}}\right]$ &$2\nu_e\sqrt{\frac{E_\alpha}{\epsilon}}$\\
  $M\geq 2E_\alpha$ & $2\nu_e$ & $\nu_e \left[ 1+ \sqrt{\frac{E_\alpha}{\epsilon}}\right]$ &-\\
  \hline
  \end{tabular}
  \label{tableDOS}
\end{table}
The energy $E$ dependence of DOS also depends on these states.
The energy dependence of the DOS in the NRM state is the same as that in the 2DEG.
The energy dependence in the RRM state is the same as that in the one-dimensional electron gas (1DEG), despite the presence of a two-dimensional system (see Table \ref{tableDOS})\cite{Jantayod13,Oshima18}:
${\rm DOS}\propto 1/\sqrt{\epsilon}$ with $\epsilon=E+E_\alpha+E_c$.
This energy dependence is irrelevant to the perpendicular magnetization.
In the ARM state, the DOS is decomposed into 1DEG and 2DEG parts\cite{Oshima18} (see Table \ref{tableDOS}).

\subsection{Method of calculation}
We consider a two-dimensional NM/FRM junction.
A gate voltage $V_g$ applied on the FRM changes position of the Fermi level in the FRM.
The interfacial barrier is assumed to be a delta function\cite{Srisongmuang08,Jantayod13,Fukumoto15} for the conservation of the $y$-component of the momentum $k_y$.
We also assume that the width of the junction along the $y$-direction, $L$, is sufficiently large.

The model Hamiltonian of the junction can be described as \cite{Rashba60,Cayao15,Streda03,Fukumoto15,Oshima18}
\begin{align}
\label{eq:total}
&H= H_{\rm NM}\theta(-x) + Z\delta(x) + (H_{\rm FRM}+eV_g)\theta(x), \\
&H_{\rm NM} = \frac{\hbar^2k^2}{2m_L},\nonumber
\end{align}
where $H_{\rm NM}$ is the Hamiltonian of the NM, and $m_L$ is the effective mass in the NM.
$\theta$ and $\delta$ are the Heaviside step function and delta function, respectively.
The expression $eV_g(\geq 0)$ is the electric potential caused by the gate voltage $V_g$, where $e(>0)$ is the elementary charge.
The expression $Z\delta(x)$ indicates the interfacial barrier, where $Z$ is the strength of the interfacial barrier.

To obtain the conductance under the low-temperature limit, we consider the wave function at the Fermi level.
The wave function in the NM, $\psi^{s}(x<0,y)$ is decomposed into the injected wave function $\psi^{s}_{\textrm{in}}$
and the reflected one $\psi^{s}_{\textrm{ref}}$ as 
\begin{align} \label{eq:left}
&\psi^{s}(x<0,y)=\psi_{\textrm{in}}^{s}+\psi_{\textrm{ref}}^{s},\\
&\psi_{\textrm{in}}^{s}=\chi_{s}e^{i(k_{F,x} x+k_y y)}\label{eq:in},\\
&\psi_{\textrm{ref}}^{s}=\left[r^{s}_\uparrow \chi_\uparrow  +r^{s}_\downarrow \chi_\downarrow \right]e^{i(-k_{F,x} x+k_y y)},
\label{eq:ref}
\end{align}
with $\chi_\uparrow = \begin{pmatrix}1 \\ 0 \end{pmatrix}$, $\chi_\downarrow = \begin{pmatrix}0 \\ 1 \end{pmatrix}$, and $k_{F,x}^2+k_y^2=k_F^2$.
Here, ($k_{F,x}$, $k_y$), $E_F$, and $k_F=\sqrt{2m_L E_F}/\hbar$ are the momentum of the injected wave, Fermi energy, and Fermi momentum, respectively.
It is noted that $s=\uparrow$, $\downarrow$ indicates the spin of an injected electron, while $r^{s}_\uparrow$ [$r^{s}_\downarrow$] denotes the reflection coefficient of up- [down-] spin electrons with up- (down-) spin injection.
The transmitted wave function $\psi^{s}(x>0,y)\equiv\psi_{\textrm{tra}}^{s}$ is shown as\cite{Oshima18} 
\begin{align}
&\psi_{\rm tra}^{s}=t_1^{s} \chi_1(\bm{k}_1)e^{i\bm{k}_1\cdot\bm{r}}+t_2^{s}\chi_-(\bm{k}_2)e^{i\bm{k}_2\cdot\bm{r}},
\label{transmissionwave}
\end{align}
with
\begin{align}
&\chi_1(\bm{k})=\theta(\Delta)\chi_+(\bm{k})+\theta(-\Delta)\chi_-(\bm{k}),\quad \Delta\equiv E-eV_g+E_c,  \\
&\chi_+(\bm{k})=\begin{pmatrix} g_+(\bm{k}) \\ 1 \end{pmatrix},\quad \chi_-(\bm{k})=\begin{pmatrix} 1 \\  g_-(\bm{k}) \end{pmatrix},\\ 
&g_\pm(\bm{k})=-\frac{\alpha i\left(k_x\mp ik_y\right)}{M+\sqrt{ \alpha^2 k^2+M^2}}\nonumber.
\end{align}
Here, $t_1^{s}$ $(t_2^{s})$ denotes the transmission coefficient with each spin injection.
The eigenfunction for $E_\pm$ is represented by $\chi_\pm$.
Here, $\bm{k}_1 = (k_{1,x},k_y )$ and $\bm{k}_2 = (k_{2,x},k_y )$ are the momenta in the FRM, which are defined for $k_1^2\leq k_2^2$ with $k_{1(2)}^2=k_{1(2),x}^2+k_y^2$.
From the energy dispersion of FRM, $k_{1(2)}^2$ is given by
\begin{align}
k_{1(2)}^2&=\frac{2m_R}{\hbar^2}\left[(E_F-eV_g)+2E_\alpha\right.\nonumber\\
&\qquad\qquad\left.-(+)\sqrt{4E_\alpha (E_F-eV_g)+4E_\alpha^2+M^2}\right].
\label{k1(2)}
\end{align}
Here, the Fermi level of the FRM is shifted by $V_g$.
The signs of $k_{1,x}$ and $k_{2,x}$ are determined so that the velocity $v_x$ takes a positive value, because the electron is injected along the $x$-direction.
The velocity operator $v_x\equiv \partial H/(\hbar\partial k_x)$ is given by\cite{Zulicke01,Streda03,Molenkamp01,Srisongmuang08,Dario15}
\begin{align}
&v_x =-\frac{i \hbar}{m(x)}\frac{\partial}{\partial x} +\frac{\alpha}{\hbar} \theta(x) \sigma_y,\label{velocity}\\
&m(x)=m_L\theta(-x)+m_R\theta(x)\nonumber
\end{align}
When $k_{1,x}$ ($k_{2,x}$) becomes an imaginary number, 
its sign is determined by $e^{i k_{1(2),x} x} \to 0$ in the limit of $x \to \infty$.

The boundary conditions at the interface \cite{Molenkamp01,Srisongmuang08,Zulicke01,Jantayod13,Fukumoto15,Reeg17} are given as follows:
\begin{align}
\begin{aligned}
&\psi^{s}(+0,y)-\psi^{s}(-0,y)=0,\\
&v_x[\psi^{s}(+0,y)-\psi^{s}(-0,y)]=\frac{2Z}{i\hbar}\psi^{s}(0,y).
\end{aligned}
\label{eq:condition}
\end{align}
From these conditions, we have
\begin{align}
&\begin{pmatrix}\chi_1 & \chi_2 & -\chi_\uparrow & -\chi_\downarrow \\ 
\left(v_{1,x}-\frac{2Z}{i\hbar}\right)\chi_1 & \left(v_{2,x}-\frac{2Z}{i\hbar}\right)\chi_2 & -v_{F,x}\chi_\uparrow & -v_{F,x}\chi_\downarrow\end{pmatrix}
\begin{pmatrix}t_1^{s}\\ t_2^{s}\\ r_\uparrow^{s}\\ r_\downarrow^{s}\end{pmatrix}\nonumber\\
&\qquad\qquad=
\begin{pmatrix}\chi_{s} \\ v_{F,x}\chi_{s}\end{pmatrix},
\label{eq:coefficient}
\end{align}
with $v_{F,x}=\hbar k_{F,x}/m_L$ and $v_{1(2),x} =\hbar k_{1(2),x}/m_R +\alpha \sigma_y/\hbar$.
Solving the coefficient from Eq. (\ref{eq:coefficient}), we obtain the $k_y$-resolved transmission probability $T^{s}(k_y)$ as\cite{Zulicke01,Srisongmuang08,Jantayod13,Oshima18} 
\begin{align}
T^{s}(k_y)&=1-R^{s}(k_y) \equiv 1-{\rm Re}\left|\frac{\psi_{\rm ref}^{s\dagger} v_x\psi^{s}_{\rm ref}}{\psi_{\rm in}^{s\dagger} v_x\psi^{s}_{\rm in}}\right| \nonumber\\
&=1-\left({|r^{s}_\uparrow|}^2+{|r^{s}_\downarrow|}^2\right).
\label{T1}
\end{align}
Here, we use the relation $R^{s}(k_y)+T^{s}(k_y)=1$, where $R^{s}(k_y)$ is the $k_y$-resolved reflection probability.

Using the Landauer formula at the low-temperature limit, electric current $I$ from the left lead to the right lead is given as follows \cite{Sablikov07,Srisongmuang08,Jantayod13,Oshima18}:
\begin{align}
I&= \frac{e^2VL}{4\pi^2\hbar}\int_{-k_F}^{k_F} dk_y[T^{\uparrow}(k_y)+T^{\downarrow}(k_y)]\nonumber\\
&= VL\frac{e^2}{2\pi\hbar}[\mathcal{T}^\uparrow+\mathcal{T}^\downarrow]=VL\mathcal{G},\label{current}
\end{align}
with
\begin{align}
&\mathcal{T}^{s(=\uparrow,\downarrow)}\equiv \frac{1}{2\pi}\int^{k_F}_{-k_F}dk_y T^{s}(k_y),\label{def:T}\\
&\mathcal{G}\equiv \frac{e^2}{2\pi\hbar}[\mathcal{T}^\uparrow+\mathcal{T}^\downarrow],\label{ddef:G}
\end{align}
where $V$ is the bias voltage, and $\mathcal{T}^{s}$ is the transmission probability.
We define the charge conductance as $dI/(LdV)$, and the charge conductance equals $\mathcal{G}$.
Hereafter, we set $\alpha >0$, $M\geq 0$, and $m_L=m_R\equiv m$ for simplicity.

\section{Results and discussion}
\label{sec:III}
\begin{figure}[htbp]
  \centering
  \includegraphics[width = 65mm]{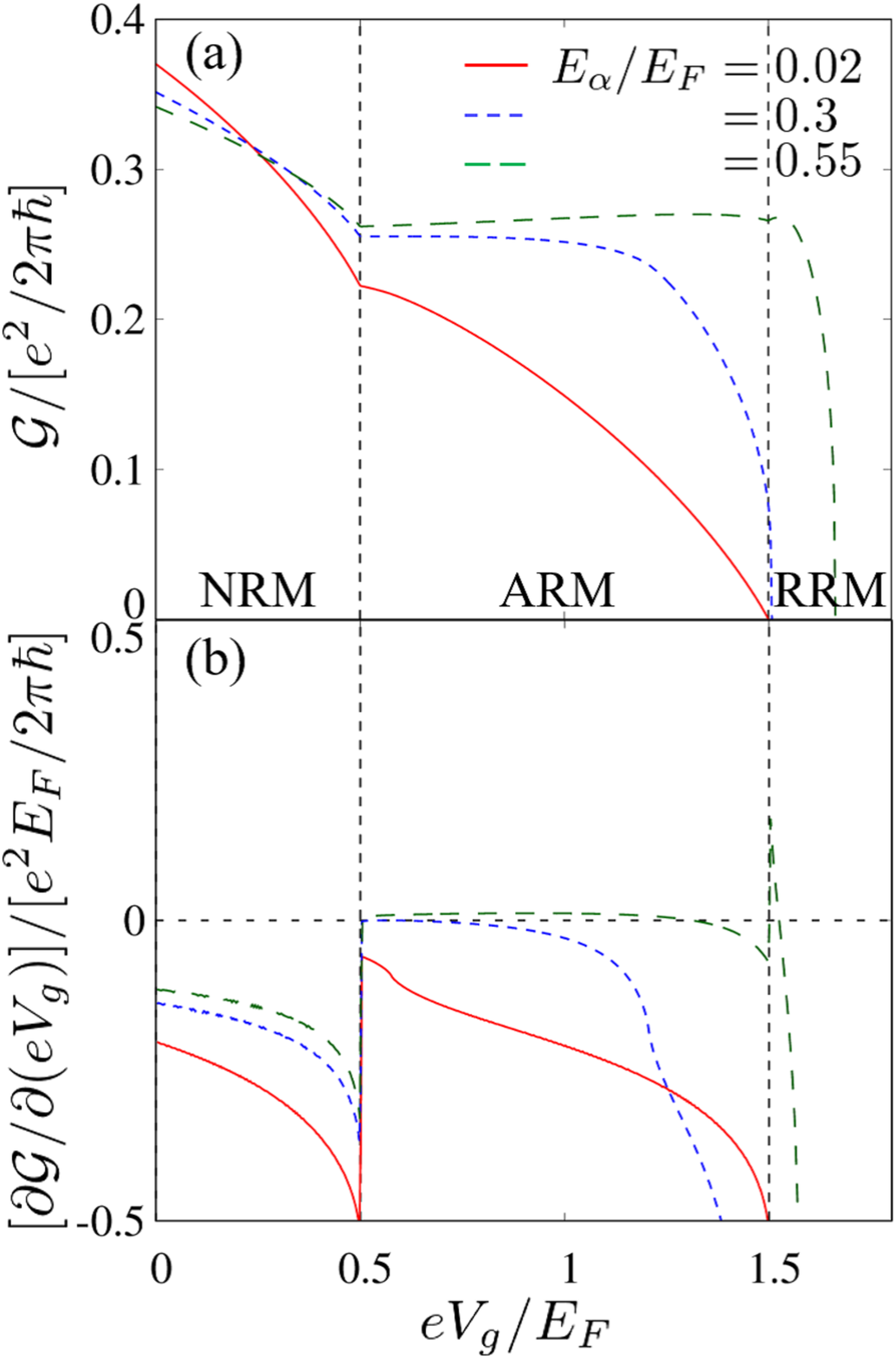}
  \caption{(Color online) $V_g$ dependence of (a) the conductance $\mathcal{G}$ and (b) $\partial\mathcal{G}/\partial (eV_g)$ in the NM/FRM junction for several values of $E_\alpha$. Here, we set $M/E_F=0.5$ and $Zk_F/E_F=1.0$, where $Z$ is the strength of the interfacial barrier, and $e(>0)$ is the elementally charge.
  For increasing $V_g$, the position of the Fermi level in the FRM approaches the band bottom of the FRM.
  We plot (b) in $-0.5 \leq [\partial\mathcal{G}/\partial (eV_g)]/[e^2 E_F/2\pi\hbar]\leq 0.5$.}
\label{add3}
\end{figure}
First, we explain the sign change of $\partial\mathcal{G}/\partial (eV_g)$ in the NM/FRM junction, as obtained in the previous paper\cite{Oshima18}.
We found that the $V_g$ dependence of $\mathcal{G}$ depends on the states of the FRM.
In the NRM state, $\mathcal{G}$ monotonically decreases with increasing $V_g$, i.e., $\partial\mathcal{G}/\partial (eV_g)<0$.
This $\partial\mathcal{G}/\partial (eV_g)<0$ can be understood by decreasing the group velocity in the FRM.
This is the conventional $V_g$ dependence of charge conductance.
However, for strong RSOC, the sign of $\partial\mathcal{G}/\partial (eV_g)$ changes at the boundary between the ARM and RRM states (see $E_\alpha/E_F=0.55$ in Fig. \ref{add3}).

\begin{figure}[htbp]
  \centering
  \includegraphics[width = 65mm]{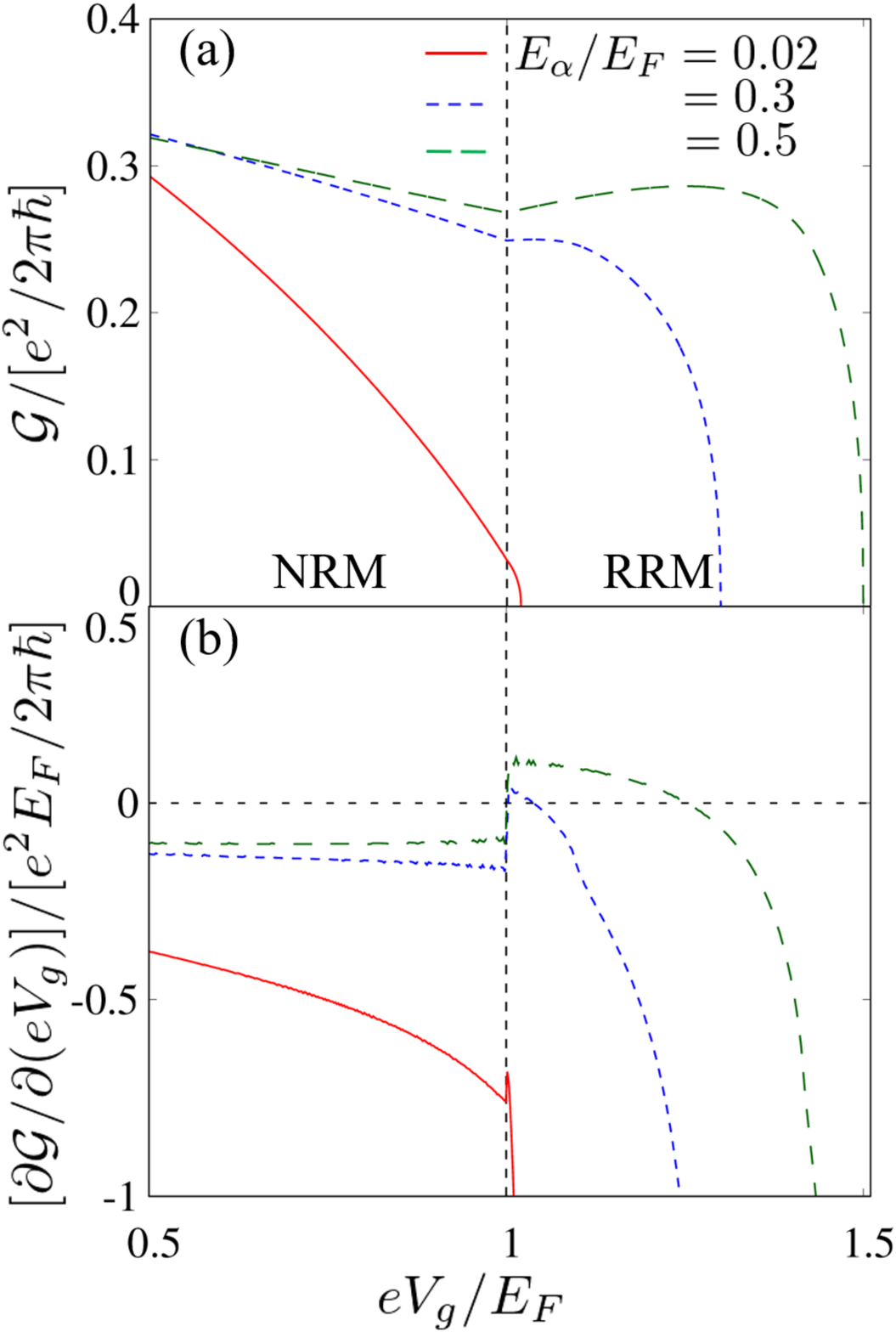}
  \caption{(Color online) $V_g$ dependence of (a) $\mathcal{G}$ and (b) $\partial\mathcal{G}/\partial(eV_g)$ with $M\to 0$ for several values of $E_\alpha$.
  We set $Zk_F/E_F=1.0$.
  Here, we plot (b) in $-1 \leq [\partial\mathcal{G}/\partial (eV_g)]/[e^2 E_F/2\pi\hbar]\leq 0.5$
  }
  \label{add11}
\end{figure}
We also find that the sign change of $\partial\mathcal{G}/\partial (eV_g)$ does not require magnetization of the FRM.
We consider $\mathcal{G}$ in the NM/FRM junction with the $M\to 0$ limit.
Figure \ref{add11}(a) shows the $V_g$ dependence of $\mathcal{G}$ for strong RSOC ($E_\alpha/E_F=0.3$ and $0.5$).
It is found that the sign of $\partial\mathcal{G}/\partial (eV_g)$ changes at $eV_g/E_F=1.0$, unlike that for weak RSOC ($E_\alpha/E_F=0.02$), where $eV_g/E_F=1.0$ corresponds to the boundary between the NRM and RRM states [see Fig. \ref{add11}(b)].
\begin{figure}[hbtp]
  \centering
  \includegraphics[width = 65mm]{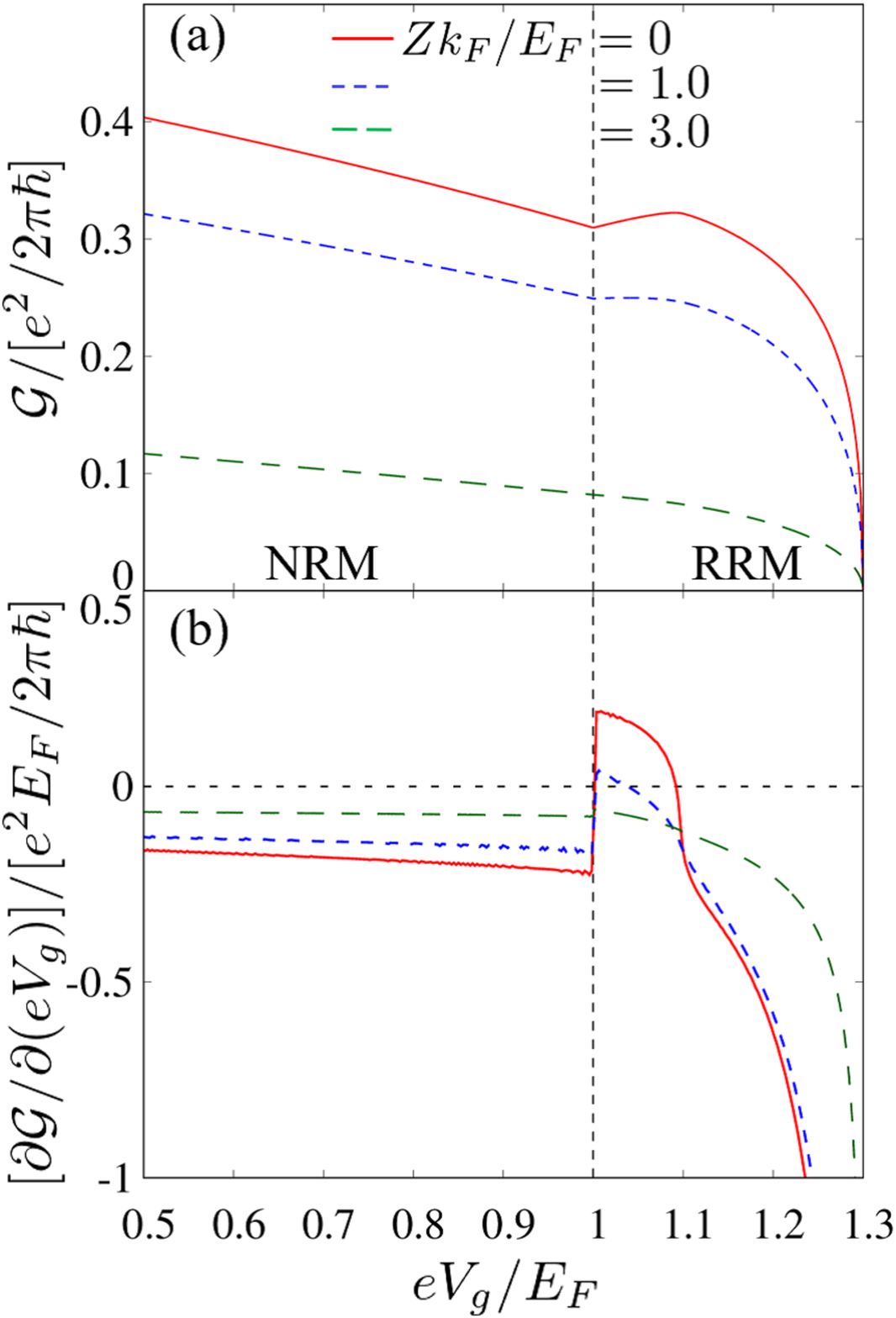}
  \caption{(Color online) $V_g$ dependence of (a) $\mathcal{G}$ and (b) $\partial\mathcal{G}/\partial(eV_g)$ with $M\to 0$ for several values of $Z$.
  We set $E_\alpha/E_F=0.3$.
  Here, we plot (b) in $-1 \leq [\partial\mathcal{G}/\partial (eV_g)]/[e^2 E_F/2\pi\hbar]\leq 0.5$.
  }
\label{add12}
\end{figure}
Figure \ref{add12}(a) shows the $V_g$ dependence of $\mathcal{G}$ for different values of $Z$ in the presence of strong RSOC ($E_\alpha/E_F=0.3$).
For $Zk_F/E_F=0.0$, the sign change is more prominent, 
but for $Zk_F/E_F=3.0$, it does not appear despite strong RSOC [see Fig. \ref{add12}(b)].
Thus, such a sign change appears even in the absence of magnetization, and it also vanishes by large $Z$, such as that in the presence of magnetization\cite{Oshima18}.

\begin{figure}[htbp]
  \centering
  \includegraphics[width = 65mm]{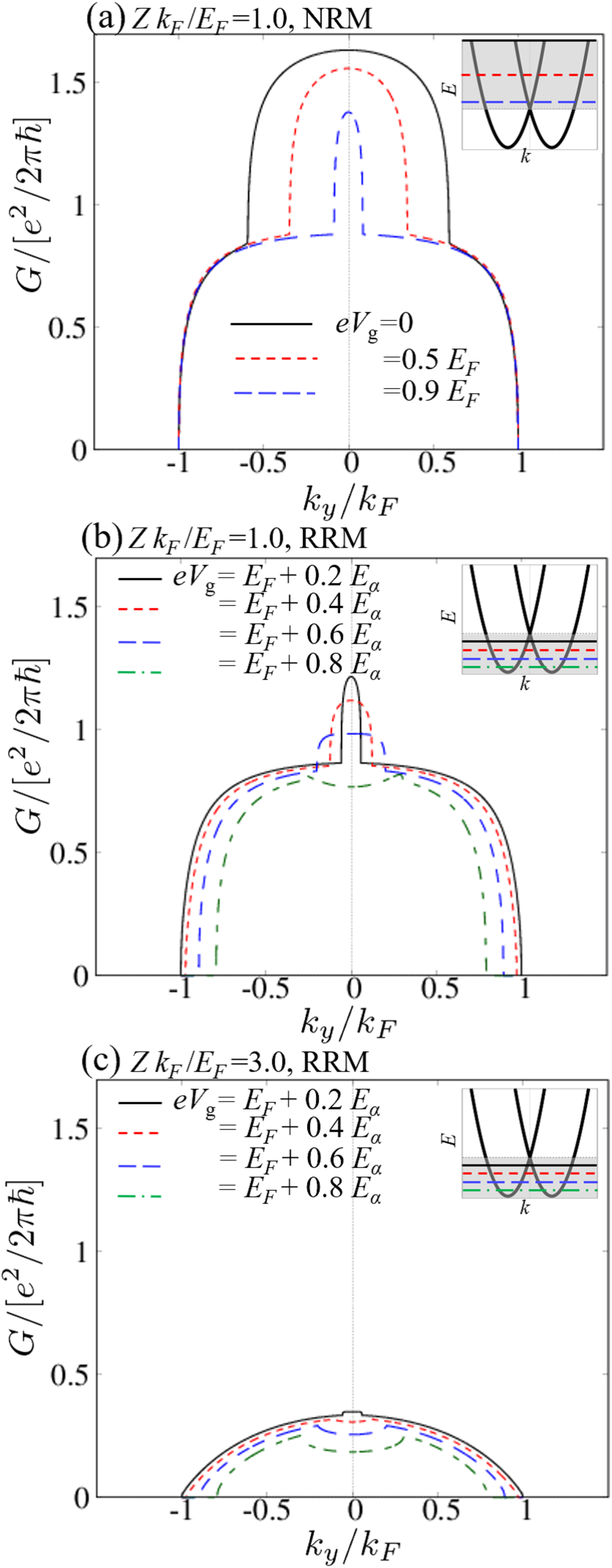}
  \caption{(Color online) $k_y$ dependence of $G$ for several $V_g$ at $E_\alpha/E_F=0.3$, and $M/E_F=0.0$.
  Here, we set (a)-(b) $Zk_F/E_F=1.0$ and (c) $Zk_F/E_F=3.0$.
  (a) and (b)-(c) correspond to the NRM and RRM states, respectively.
  The inset shows the band structure of the FRM, and the shaded region represents the region of (a) NRM and (b)-(c) RRM.}
  \label{add2}
\end{figure}

\begin{figure}[htbp]
  \centering
  \includegraphics[width = 90mm]{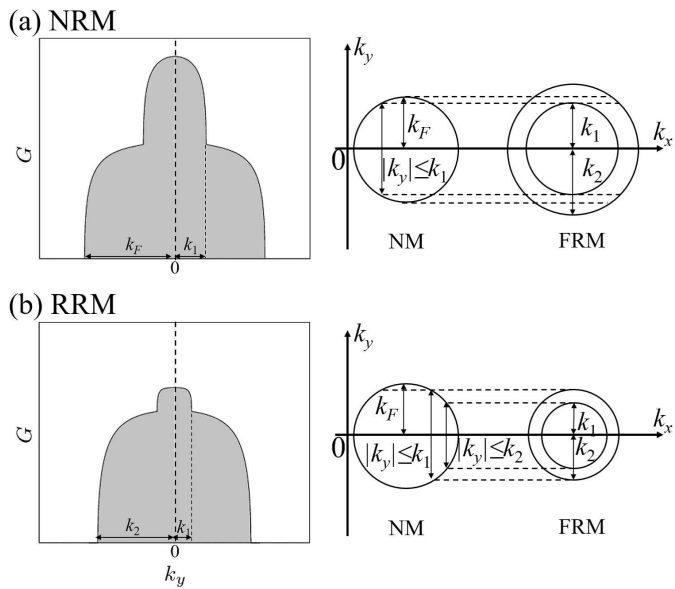}
  \caption{Schematic of the relation between $G(k_y)$ and the momentum.
  (a) NRM and (b) RRM.
  A hump in $|k_y|\leq k_1$ indicates the contribution from the inner Fermi surface of the FRM, where $k_{1(2)}$ is the momentum of the inner (outer) Fermi surface of the FRM (see right panel).
  The area of the shaded region in the left panels is proportional to $\mathcal{G}$.
  }
  \label{adddiscuss}
\end{figure}

\begin{figure}[htbp]
  \centering
  \includegraphics[width = 90mm]{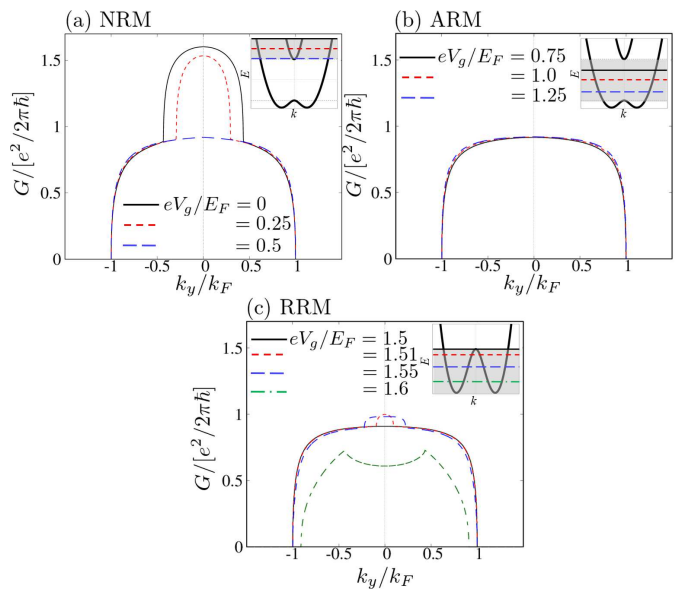}
  \caption{(Color online) $k_y$ dependence of $G$ for several $V_g$ in the NM/FRM junction at $E_\alpha/E_F=0.55$, $M/E_F=0.5$, and $Zk_F/E_F=1.0$.
  (a), (b), and (c) correspond to the NRM, ARM, and RRM states, respectively.
  The inset shows the band structure of the FRM, and its shaded region shows the region of (a) NRM, (b) ARM, and (c) RRM.}
  \label{add4}
\end{figure}

To understand the origin of the sign change of $\partial\mathcal{G}/\partial (eV_g)$,
we will show the $k_y$-resolved conductance in Figs. \ref{add2}, \ref{adddiscuss}, and \ref{add4}.
We define the $k_y$-resolved conductance $G$ as
\begin{align}
G(k_y)\equiv \frac{e^2}{2\pi\hbar}[T^\uparrow(k_y)+T^\downarrow(k_y)].
\end{align}
It is noted that the area surrounded by the line of $G(k_y)$ is proportional to $\mathcal{G}$ (see Fig. \ref{adddiscuss}), and a decrease in the area corresponds to the decrease in $\mathcal{G}$.

%
Figure \ref{add2} shows $G(k_y)$ in the (a) NRM and (b) RRM states for several values of $V_g$.
It is found that $G(k_y)$ can be decomposed into two parts: 
$G(k_y)$ in $|k_y|> k_1$ and in $|k_y|\leq k_1$, where $k_1$ is momentum of the inner Fermi surface of the FRM.
Figure \ref{add2}(a) indicates that $G(|k_y|> k_1)$ is almost independent of $V_g$.
On the other hand, $G(|k_y|\leq k_1)$ depends on $V_g$, and it takes on a particular shape, which is dubbed the hump.
The width of the hump is $2k_1$, and it decreases with increasing $V_g$.
Near the boundary between the NRM and RRM states ($eV_g/E_F\to 1.0$), its width goes to zero, but the hump exists.
It is expected that the existence of the hump reflects a band crossing of $E_\pm$ [see Fig. \ref{bandstructure}(a)].

It is noticed that $G(|k_y|\leq k_1)$ is obtained from the two wave functions, whose momentum correspond to the inner and outer Fermi surface of the FRM.
As a result, $G(|k_y|\leq k_1)$ includes the contribution from the inner and outer Fermi surface, as shown in Fig. \ref{adddiscuss}.
It is noted that for $M\neq 0$, there is no inner Fermi surface in the ARM state (we show results in $M \neq 0$ later).
Then, in the ARM state, the hump structure of $G$ vanishes, and $G$ takes a nonzero value, which is purely caused by the contribution of the outer Fermi surface.
In addition, we note that the width of the hump is suppressed by decreasing the size of the inner Fermi surface.
As a result, this hump originates from the inner Fermi surface.

In the RRM state, $G(k_y)$ is also decomposed into $G(|k_y|> k_1)$ and $G(|k_y|\leq k_1)$ [see Figs. \ref{add2}(b)-(c)].
$G(|k_y|\leq k_1)$ comes from the inner Fermi surface.
We find that $G(|k_y|> k_1)$ is almost irrelevant, but $G(|k_y|\leq k_1)$ is relevant to the sign change of $\partial\mathcal{G}/\partial (eV_g)$.
$G(|k_y|\leq k_1)$ in the RRM state has a hump, which is similar to that in the NRM state.
The height of the hump is suppressed, and the shape of the hump changes from convex upward to downward by increasing $V_g$.
It contributes to the decrease in $\mathcal{G}$.
Furthermore, the width of the hump $2k_1$ becomes wide with increasing $V_g$, and it contributes to the increase in $\mathcal{G}$.
As a result, the sign of $\partial\mathcal{G}/\partial (eV_g)$ is determined by the tradeoff between the suppression of the height of the hump and the increase in the width.
It is noted that in the presence of the large interfacial barrier, the height of the hump is suppressed, and the increase in $\mathcal{G}$ caused by increase in the width is also suppressed.
Then, the tradeoff leads $\partial\mathcal{G}/\partial (eV_g)<0$ [see Figs. \ref{add12} and \ref{add2}(b)-(c)].
In short, such a tradeoff determines the sign of $\partial\mathcal{G}/\partial (eV_g)$, and its tradeoff is given by the inner Fermi surface of FRM, which is fragile against the interfacial barrier.
In other words, the sign change of $\partial \mathcal{G}/\partial (eV_g)$ occurs when the contribution from the inner Fermi surface is sufficiently large.
Actually, $\partial\mathcal{G}/\partial (eV_g) <0$ is obtained in weak RSOC, where the size of the inner Fermi surface is sufficiently small [see Fig. \ref{add11}].

Moreover, we study the origin of the sign change of $\partial \mathcal{G}/\partial V_g$ in nonzero $M$.
In the presence of $M$, the ARM state additionally appears, where there is no inner Fermi surface in the FRM [see Figs. \ref{bandstructure}(b)-(c)].
As a result, the hump vanishes at the boundary between the NRM and ARM states, unlike that in the $M\to 0$ limit [see $eV_g=0.5 E_F$ in Fig. \ref{add4}(a)].
Furthermore, the hump disappears in the ARM state, as shown in Fig. \ref{add4}(b).
It is noticed that this $G(k_y)$ is almost independent of $V_g$ in the ARM state.
However, in the RRM state, $G(k_y)$ depends on $V_g$, and the hump is reproduced in $|k_y|\leq k_1$.
After the hump becomes a certain size, $G(|k_y|\leq k_1)$ collapses with increasing $V_g$, like that in the RRM state in the $M\to0$ limit.
Similar to the charge conductance $\mathcal{G}$ for the $M \to 0$ limit, the reproduction of the hump in $|k_y|\leq k_1$ contributes to the sign change of $\partial\mathcal{G}/\partial V_g$, and the existence of the hump indicates the contribution of the inner Fermi surface.
Thus, even for nonzero $M$, the origin of the sign change of $\partial\mathcal{G}/\partial(eV_g)$ also stems from the inner Fermi surface.

In Figs. \ref{add2}(b)-(c) and \ref{add4}(c), we notice that the momentum range of $G(k_y)\neq 0$ changes from $2k_F$ to $2k_2$, and the value of $2k_2$ decreases with increasing $V_g$.
The decrease in the range appears because the size of the outer Fermi surface of the FRM becomes smaller than that of the NM,
reducing the magnitude of $\mathcal{G}$.
Then, $\partial\mathcal{G}/\partial(eV_g)$ becomes negative even if the tradeoff of the inner Fermi surface contributes to increasing $\mathcal{G}$ when the decrease in $\mathcal{G}$ caused by the decrease of $2k_2$ is sufficiently large.

Finally, we discuss whether the charge conductance depends on the spin structure of the RSOC.
In this study, we consider the charge conductance only for $\alpha>0$.
However, the obtained charge conductance is independent of the sign of $\alpha$.
Because the sign of $\alpha$ refers to the helicity of the spin texture of the RSOC, the obtained charge conductance is independent of the helicity.
Furthermore, under the transformation $\sigma_x(\sigma_y)\to \sigma_y(\sigma_x)$ [i.e., RSOC $\to$ linear Dresselhaus type spin-orbit coupling (DSOC) (Hamiltonian of DSOC: $H_{\rm DSOC}=\alpha (k_x\sigma_x-k_y\sigma_y$))]\cite{Dresselhaus55}, the obtained conductance is invariant (see \ref{app3}).
As a result, we expect that the charge conductance is independent of the spin texture.

\section{Conclusion}
\label{sec:V}
In this study, we calculated the gate voltage ($V_g$) dependence of charge conductance in the NM/FRM junction.
Additionally, we clarified the physical origin of the unconventional $V_{g}$ dependence of charge conductance in NM/FRM junction, where charge conductance increases with $V_{g}$.
The origin of the unconventional $V_{g}$ dependence is due to the non-monotonic change in the size of the inner Fermi surface in FRM as a function of $V_{g}$.
This result does not change by the spin transformation $\sigma_x(\sigma_y)\to \sigma_y(\sigma_x)$, in which the RSOC replaces linear DSOC.

Furthermore, we studied the charge transport of the FRM.
There are several future areas of study.
For example, because the FRM has specific spin structures with non-zero Berry curvature, 
we can expect new ballistic transport phenomena, unlike that in diffusive transport, such as Edelstein effect \cite{Taguchi17}.
In addition, the calculation of the tunneling magnetoresistance (TMR) in FRM junctions will be interesting; TMR may show different features depending on the gate voltage.

\section*{Acknowledgment}
This work was supported
by a Grant-in-Aid for Scientific Research on Innovative
Areas, Topological Material Science (Grants No. JP15H05851,
No. JP15H05853 No. JP15K21717), a Grant-in-Aid
for Challenging Exploratory Research (Grant No. JP15K13498)
from the Ministry of Education, Culture, Sports, Science, and
Technology, Japan (MEXT), the Core Research for
Evolutional Science and Technology (CREST) of the Japan
Science and Technology Corporation (JST) (Grant No. JPMJCR14F1).

\appendix
\section{Spin-dependent Conductance under the Unitary Transformation}\label{app3}
In this appendix, we show that the obtained transmission probabilities are independent of the detail of the spin texture caused by RSOC. First, we show that the conductance is invariant under unitary transformation $U$, which satisfies $\acute{v}_x= Uv_xU^\dagger$.
Here, $\acute{v}_x$ is a velocity operator after transformation $U$.
Applying $U$, the Hamiltonian of the junction [see Eq. (\ref{eq:total})] and the eigenfunctions on each side are changed as $H\to \acute{H}= UHU^\dagger$ and $\chi_{\uparrow(\downarrow,1,2)}\to\acute{\chi}_{\uparrow(\downarrow,1,2)}= U\chi_{\uparrow(\downarrow,1,2)}$.
From the boundary condition after the transformation, the transmission and reflection coefficients, $\acute{t}^{s}_{1[2]}$ and $\acute{r}^{s}_{\uparrow[\downarrow]}$, are given by
\begin{align}
&\begin{pmatrix}U&0 \\ 0&U\end{pmatrix}\begin{pmatrix}\chi_1 & \chi_2 & -\chi_\uparrow & -\chi_\downarrow \\
\left(v_{1,x}-\frac{2Z}{i\hbar}\right)\chi_1 & \left(v_{2,x}-\frac{2Z}{i\hbar}\right)\chi_2 & -v_{F,x}\chi_\uparrow & -v_{F,x}\chi_\downarrow\end{pmatrix}\nonumber\\
&\qquad\cdot\begin{pmatrix}\acute{t}_1^{s}\\ \acute{t}_2^{s}\\ \acute{r}_\uparrow^{s}\\ \acute{r}_\downarrow^{s}\end{pmatrix}
=
\begin{pmatrix}U&0 \\ 0&U\end{pmatrix}\begin{pmatrix}\chi_{s} \\ v_{F,x}\chi_{s}\end{pmatrix}.
\label{eq:coefficientdash}
\end{align}
$U$ is a $2\times 2$ matrix and $\chi_{\uparrow(\downarrow,1,2)}$ are two-component vectors.
Here, we notice that Eq. (\ref{eq:coefficientdash}) is the same as Eq. (\ref{eq:coefficient}),
i.e., $\acute{t}^{s}_{1[2]}=t^{s}_{1[2]}$ and $\acute{r}^{s}_{\uparrow[\downarrow]}=r^{s}_{\uparrow[\downarrow]}$.
Then, we obtain $\acute{\psi}^{s}_{\rm in[ref,tra]}=U\psi^{s}_{\rm in[ref,tra]}$, where $\acute{\psi}^{s}_{\rm in[ref,tra]}$ is the wave function after the transformation.
As a result, we find that the transmission probabilities are invariant about the transformation $U$:
\begin{align}
{\rm Re}\left|\frac{\psi_{\rm tra}^{s\dagger} v_x\psi^{s}_{\rm tra}}{\psi_{\rm in}^{s\dagger} v_x\psi^{s}_{\rm in}}\right|
&={\rm Re}\left|\frac{\acute{\psi}_{\rm tra}^{s\dagger}U v_x U^\dagger \acute{\psi}^{s}_{\rm tra}}{\acute{\psi}_{\rm in}^{s\dagger}U v_x U^\dagger \acute{\psi}^{s}_{\rm in}}\right|\nonumber\\
&={\rm Re}\left|\frac{\acute{\psi}_{\rm tra}^{s\dagger}\acute{v}_x \acute{\psi}^{s}_{\rm tra}}{\acute{\psi}_{\rm in}^{s\dagger}\acute{v}_x \acute{\psi}^{s}_{\rm in}}\right|.
\end{align}

An NM/NM with RSOC junction can be connected to an NM/NM with DSOC junction through a unitary transformation $R=i(\sigma_x+\sigma_y)/\sqrt{2}$\cite{Lucignano08,Dario15}, which changes as $\sigma_x \to \sigma_y$, $\sigma_y \to \sigma_x$, and $\sigma_z \to -\sigma_z$.
The NM/FRM junction corresponds to NM/FM+DSOC junction through the transformation $R$ and $P$.
$P=V_{k_y}\sigma_x$ is the spin and momentum transformation, which changes $k_y\to -k_y$, $\sigma_y\to -\sigma_y$, and $\sigma_z\to-\sigma_z$.
Here, $V_{k_y}$ is a mirror operator, which changes as $(k_x,k_y,k_z)\to (k_x,-k_y,k_z)$.
Then, the transmission probabilities in the RSOC are equal to that in the DSOC because $R$ and $P$ satisfy our assumption.
These junctions are different from each other only in terms of the spin texture caused by the SOC.
Therefore, it is expected that these transmission probabilities can be understood without considering the spin texture, which is caused by the SOC.
It is noted that $k_y$-resolved transmission probabilities in RSOC correspond to that in DSOC with $k_y\to -k_y$ because of transformation $P$.


\begin{thebibliography}{10}

  \bibitem{Rashba60}
  Emmanuel~I Rashba.
  \newblock Properties of semiconductors with an extremum loop. 1. cyclotron and
    combinational resonance in a magnetic field perpendicular to the plane of the
    loop.
  \newblock {\em Sov. Phys. Solid State}, 2(6):1109--1122, 1960.
  
  \bibitem{Datta90}
  Supriyo Datta and Biswajit Das.
  \newblock Electronic analog of the electro-optic modulator.
  \newblock {\em Appl. Phys. Lett.}, 56(7):665--667, 1990.
  
  \bibitem{Nitta97}
  Junsaku Nitta, Tatsushi Akazaki, Hideaki Takayanagi, and Takatomo Enoki.
  \newblock Gate control of spin-orbit interaction in an inverted
    i${\mathrm{n}}_{0.53}$g${\mathrm{a}}_{0.47}$as/i${\mathrm{n}}_{0.52}$a${\mathrm{l}}_{0.48}$as
    heterostructure.
  \newblock {\em Phys. Rev. Lett.}, 78:1335--1338, Feb 1997.
  
  \bibitem{Molenkamp01}
  Laurens~W Molenkamp, Georg Schmidt, and Gerrit~EW Bauer.
  \newblock Rashba hamiltonian and electron transport.
  \newblock {\em Phys. Rev. B}, 64(12):121202, 2001.
  
  \bibitem{Jiang03}
  Y~Jiang and MBA Jalil.
  \newblock Enhanced spin injection and magnetoconductance by controllable rashba
    coupling in a ferromagnet/two-dimensional electron gas structure.
  \newblock {\em J. Phys: Cond. Mat.}, 15(2):L31, 2003.
  
  \bibitem{Cai08}
  Lei Cai, YC~Tao, Jing-guo Hu, and Guo-jun Jin.
  \newblock Effect of rashba spin--orbit coupling on the spin-polarized transport
    in ferromagnet/semiconductor double tunnel junctions.
  \newblock {\em Physics Letters A}, 372(32):5361--5367, 2008.
  
  \bibitem{Jiang10}
  Kai-Ming Jiang, Rong Zhang, Jun Yang, Chun-Xiao Yue, and Zu-Yao Sun.
  \newblock Tunneling magnetoresistance properties in ballistic spin field-effect
    transistors.
  \newblock {\em IEEE Transactions on Electron Devices}, 57(8):2005--2012, 2010.
  
  \bibitem{Streda03}
  P~St{\v{r}}eda and P~{\v{S}}eba.
  \newblock Antisymmetric spin filtering in one-dimensional electron systems with
    uniform spin-orbit coupling.
  \newblock {\em Phys. Rev. Lett.}, 90(25):256601, 2003.
  
  \bibitem{Srisongmuang08}
  B~Srisongmuang, P~Pairor, and M~Berciu.
  \newblock Tunneling conductance of a two-dimensional electron gas with rashba
    spin-orbit coupling.
  \newblock {\em Phys. Rev. B}, 78(15):155317, 2008.
  
  \bibitem{Tang12}
  Chi-Shung Tang, Shu-Yu Chang, and Shun-Jen Cheng.
  \newblock Finger-gate manipulated quantum transport in a semiconductor narrow
    constriction with spin-orbit interactions and zeeman effect.
  \newblock {\em Phys. Rev. B}, 86(12):125321, 2012.
  
  \bibitem{Jantayod13}
  A~Jantayod and P~Pairor.
  \newblock Charge and spin transport across two-dimensional non-centrosymmetric
    semiconductor/metal interface.
  \newblock {\em Phys. E: Low-dimensional Systems and Nanostructures},
    48:111--117, 2013.
  
  \bibitem{Rainis14}
  Diego Rainis and Daniel Loss.
  \newblock Conductance behavior in nanowires with spin-orbit interaction: A
    numerical study.
  \newblock {\em Phys. Rev. B}, 90(23):235415, 2014.
  
  \bibitem{Jantayod15}
  A~Jantayod and P~Pairor.
  \newblock Tunneling conductance of a metal/two-dimensional electron gas with
    rashba spin--orbit coupling junction within a lattice model.
  \newblock {\em Superlattices and Microstructures}, 88:541--550, 2015.
  
  \bibitem{Cayao15}
  Jorge Cayao, Elsa Prada, Pablo San-Jose, and Ram{\'o}n Aguado.
  \newblock Sns junctions in nanowires with spin-orbit coupling: Role of
    confinement and helicity on the subgap spectrum.
  \newblock {\em Phys. Rev. B}, 91(2):024514, 2015.
  
  \bibitem{Tang17}
  Chi-Shung Tang, Yun-Hsuan Yu, Nzar~Rauf Abdullah, and Vidar Gudmundsson.
  \newblock Transport signatures of top-gate bound states with strong
    rashba-zeeman effect.
  \newblock {\em Physics Letters A}, 381(47):3960 -- 3963, 2017.
  
  \bibitem{Oshima18}
  Daisuke Oshima, Katsuhisa Taguchi, and Yukio Tanaka.
  \newblock Tunneling conductance in two-dimensional junctions between a normal
    metal and a ferromagnetic rashba metal.
  \newblock {\em Journal of the Physical Society of Japan}, 87(3):034710, 2018.
  
  \bibitem{Fukumoto15}
  Toshiyuki Fukumoto, Katsuhisa Taguchi, Shingo Kobayashi, and Yukio Tanaka.
  \newblock Theory of tunneling conductance of anomalous rashba
    metal/superconductor junctions.
  \newblock {\em Phys. Rev. B}, 92(14):144514, 2015.
  
  \bibitem{Ganichev14}
  Sergey~D. Ganichev and Leonid~E. Golub.
  \newblock Interplay of rashba/dresselhaus spin splittings probed by
    photogalvanic spectroscopy ?a review.
  \newblock {\em physica status solidi (b)}, 251(9):1801--1823, 2014.
  
  \bibitem{Kohda17}
  Makoto Kohda and Gian Salis.
  \newblock {Physics and application of persistent spin helix state in
    semiconductor heterostructures}.
  \newblock {\em Semicond. Sci. Technol}, 32, 2017.
  
  \bibitem{Ast07}
  Christian~R Ast, J{\"u}rgen Henk, Arthur Ernst, Luca Moreschini, Mihaela~C
    Falub, Daniela Pacil{\'e}, Patrick Bruno, Klaus Kern, and Marco Grioni.
  \newblock Giant spin splitting through surface alloying.
  \newblock {\em Phys. Rev. Lett.}, 98(18):186807, 2007.
  
  \bibitem{Sablikov07}
  Vladimir~A. Sablikov and Yurii~Ya. Tkach.
  \newblock Evanescent states in two-dimensional electron systems with spin-orbit
    interaction and spin-dependent transmission through a barrier.
  \newblock {\em Phys. Rev. B}, 76:245321, Dec 2007.
  
  \bibitem{Ast08}
  Christian~R Ast, Daniela Pacil{\'e}, Luca Moreschini, Mihaela~C Falub, Marco
    Papagno, Klaus Kern, Marco Grioni, J{\"u}rgen Henk, Arthur Ernst, Sergey
    Ostanin, and Patrick Bruno.
  \newblock Spin-orbit split two-dimensional electron gas with tunable rashba and
    fermi energy.
  \newblock {\em Phys. Rev. B}, 77(8):081407, 2008.
  
  \bibitem{Mathias10}
  S~Mathias, A~Ruffing, F~Deicke, M~Wiesenmayer, I~Sakar, G~Bihlmayer,
    EV~Chulkov, Yu~M Koroteev, PM~Echenique, M~Bauer, and M~Aeschlimann.
  \newblock Quantum-well-induced giant spin-orbit splitting.
  \newblock {\em Phys. Rev. Lett.}, 104(6):066802, 2010.
  
  \bibitem{Ishizaka11}
  K~Ishizaka, MS~Bahramy, H~Murakawa, M~Sakano, T~Shimojima, T~Sonobe, K~Koizumi,
    S~Shin, H~Miyahara, A~Kimura, K~Miyamoto, T~Okuda, H~Namatame, M~Taniguchi,
    R~Arita, N~Nagaosa, K~Kobayashi, Y~Murakami, R~Kumai, Y~Kaneko, Y~Onose, and
    Y~Tokura.
  \newblock Giant rashba-type spin splitting in bulk bitei.
  \newblock {\em Nat. Mater}, 10(7):521--526, 2011.
  
  \bibitem{Zulicke01}
  U~Z{\"u}licke and C~Schroll.
  \newblock Interface conductance of ballistic ferromagnetic-metal-2deg hybrid
    systems with rashba spin-orbit coupling.
  \newblock {\em Phys. Rev. Lett.}, 88(2):029701, 2001.
  
  \bibitem{Dario15}
  Dario Bercioux and Procolo Lucignano.
  \newblock {Quantum transport in Rashba spin^^e2^^80^^93orbit materials: a
    review}.
  \newblock {\em Rep. Prog. Phys}, 78, 2015.
  
  \bibitem{Reeg17}
  Christopher Reeg and Dmitrii~L Maslov.
  \newblock Transport signatures of topological superconductivity in a
    proximity-coupled nanowire.
  \newblock {\em Phys. Rev. B}, 95(20):205439, 2017.
  
  \bibitem{Dresselhaus55}
  G.~Dresselhaus.
  \newblock Spin-orbit coupling effects in zinc blende structures.
  \newblock {\em Phys. Rev.}, 100:580--586, Oct 1955.
  
  \bibitem{Taguchi17}
  K.~Taguchi, B.~T. Zhou, Y.~Kawaguchi, Y.~Tanaka, and K.~T. Law.
  \newblock Valley edelstein effect in monolayer transition-metal
    dichalcogenides.
  \newblock {\em Phys. Rev. B}, 98:035435, Jul 2018.
  
  \bibitem{Lucignano08}
  P.~Lucignano, R.~Raimondi, and A.~Tagliacozzo.
  \newblock Spin hall effect in a two-dimensional electron gas in the presence of
    a magnetic field.
  \newblock {\em Phys. Rev. B}, 78:035336, Jul 2008.
  
  \end{thebibliography}
\end{document}